# A Low-Cost Natural Gas/Freshwater Aerial Pipeline

Alexander A. Bolonkin•
C & R
1310 Avenue R, Suite 6-F
Brooklyn, New York 11229, USA
aBolonkin@juno.com

Richard B. Cathcart
Geographos
1300 West Olive Avenue, Suite M
Burbank, California 91506, USA

## ABSTRACT

Offered is a new type of low-cost aerial pipeline for delivery of natural gas, an important industrial and residential fuel, and freshwater as well as other payloads over long distances. The offered pipeline dramatically decreases the construction and operation costs and the time necessary for pipeline construction. A dual-use type of freight pipeline can improve an arid rural environment landscape and provide a reliable energy supply for cities. Our aerial pipeline is a large, self-lofting flexible tube disposed at high altitude. Presently, the term "natural gas" lacks a precise technical definition, but the main components of natural gas are methane, which has a specific weight less than air. A lift force of one cubic meter of methane equals approximately 0.5 kg. The lightweight film flexible pipeline can be located in the Earth-atmosphere at high altitude and poses no threat to airplanes or the local environment. The authors also suggest using lift force of this pipeline in tandem with wing devices for cheap shipment of a various payloads (oil, coal and water) over long distances. The article contains a computed macroproject in northwest China for delivery of 24 billion cubic meter of gas and 23 millions tonnes of water annually.

## Introduction

Natural gas is a mixture of flammable gases, mainly the hydrocarbons methane ($CH_4$) and ethane that found in bulk beneath the Earth's surface. Helium is also found in relatively high concentrations in natural gas. Natural gas usually occurs in close association with petroleum. Although many natural gases can be used directly from the well without treatment, some must be processed to remove such undesirable constituents as carbon dioxide, poisonous hydrogen sulfide, and other sulfur components.

Methods of pipeline transportation that were developed in the 1920s marked a significant stage in the use of gas. After World War II there occurred a period of tremendous expansion that has continued into the 21$^{st}$ Century. Increasingly, this expansion relies on the use of pipeline transportation of gas. Among the largest accumulations of natural gas are those of Urengoy in Siberia, the Texas Panhandle in the United States, Slochteren-Groningen area in The Netherlands, and Hassi RMel in Algeria. Gas accumulations are mostly encountered in the deeper parts of sedimentary basins. Natural gas fields are often located far from the major centers of consumption. Consequently, the gas must be transported.

---

• Corresponding author.

Transportation of natural gas depends upon its form. In a gaseous form it is transported by pipeline under high pressure, and in a liquid form it is transported by tanker ship [1].

Large gas pipelines enable gas to be transported over great distances. Examples are the North American pipelines, which extend from Texas and Louisiana to the Northeast coast, and from the Alberta fields to the Atlantic seaboard. Transportation pressure is generally 70 kilograms per square centimeter (up to 200 atm) because transportation costs are lowest for pressures in this range. Natural gas pipeline diameters for such long-distance transportation have tended to increase from an average of about 60 to 70 centimeters in 1960 to about 1.20 meters nowadays. Some macroprojects involve diameters of more than 2 meters. Because of pressure losses, the pressure is boosted every 80 or 100 kilometers to keep a constant rate of flow.

Petroleum prospecting has revealed the presence of large gas fields in Africa, the Middle East, Alaska, and China. Gas is transported from developed regions by special LNG ships. The gas is liquefied to – 160 C and transported in tankers with insulated containers. Since 1965 the capacity of tankers has risen to as much as 120,000 cubic meters, which enables some tankers to convey as much as 70 million cubic meters of gas per voyage. Land or sea-based storage of low-temperature liquefied gas requires double-walled tanks with special insulation. Such tanks may hold as much as 50,000 cubic meters. Even larger storage facilities have been created by using depleted subsurface oil or gas geological reservoirs near consumption centers or by the creation of artificial gas fields in aquifer layers. The latter technique developed rapidly, and the number of storage facilities of this type in the USA increased tremendously during the late 20$^{th}$ Century. There are also such underground storage areas in France and Germany.

Residential and commercial use consumes the largest proportion of natural gas in North America and Western Europe, while industry consumes the next largest amount and electric-power generation is third in worldwide natural-gas consumption. By far the major use of natural gas is as fuel, though increasing amounts are used by the chemical industry for raw material. Among the industries that consume large volumes are food, paper, chemicals, petroleum refining, and primary metals. In the USA, a large amount fuels household heaters; in Russia a considerable volume goes for electric-power generation and to generate export revenue. Exportation and importation of natural gas involves some aspects of geopolitical assessment [2]

Most materials that can be moved in large quantities in the form of liquids, gases, or slurries (fine particles suspended in liquid) are generally moved through freight pipelines [3]. Pipelines are lines of pipe equipped with pumps, valves, and other control devices for transporting materials from their remote sources to storage tanks or refineries and in turn to distribution facilities; pipelines may also convey industrial waste and sewage to processing plants for treatment before disposal.

Pipelines vary in diameter from tiny pipes up to lines 9 meters across used in high-volume freshwater distribution and sewage collection networks. Pipelines usually consist of sections of pipe made of steel, cast iron, or aluminum, though some are constructed of concrete, fired clay products, and occasionally plastics. The sections are joined together and, in most cases, installed underground. Because great quantities of often expensive and sometimes environmentally harmful materials are carried by pipelines, it is essential that the systems be well constructed and monitored in order to ensure that they will operate smoothly, efficiently, and safely. Pipes are often covered with a protective coating of coal-tar enamel, asphalt, or plastic; sometimes these coatings may be reinforced or supplemented by an additional sheath of asbestos felt, fiberglass or polyurethane. The materials used depend on the substance to be carried and its chemical activity and possible corrosive action on the pipe. Pipeline designers must also consider such factors as the capacity of the pipeline, internal and external pressures affecting the pipeline, water- and air-tightness, and construction and operating costs. Generally the first step in construction is to clear the ground and dig a trench deep enough to allow for approximately 51 centimeters of soil overburden to cover the pipe. Sections of pipe are then held over the trench, where they are joined together by welding, riveting or mechanical coupling, covered with a protective coating, and lowered into position. Pipelines of some water-supply systems may follow the slope of the land, winding through irregular landscapes like low-gradient railroads and highways do, and rely on gravity to keep the water flowing through them. If necessary, the gravity flow is supplemented by pumping. Most pipelines, however, are operated under pressure to overcome friction within the pipe and differences in elevation. Such systems have a series of pumping stations that are located at intervals of from 80 to 320 kilometers. Many pipelines are equipped with a system of valves that may be shut in the event of a breach in the line. Nevertheless, a short-period breach could still result in a spill of oil or escapes of gas. Vigilant ground and air inspection crews help to avert such damaging and costly accidents by checking periodically the pipeline for obvious weaknesses and stresses. Various methods are used to control corrosion in metallic pipelines. It is worth noting that metallic pipelines, especially those located on the Earth's surface, are subject to Space Weather, just like electric power grids [4]. In cathodic protection, a negative electrical charge is maintained throughout the pipe to inhibit the electrochemical process of corrosion. In other cases the interior is lined with paints and coatings of plastic and rubber or wrappings of fiberglass, asphalt, or felt. Sometimes corrosion-inhibiting chemicals are injected into the cargo. Pipelines are also cleaned by passing devices called pigs; a pig may be a ball of the same diameter as the pipe; this kind of pig works by scraping clean the pipe's interior as it is propelled along by the flowing cargo. It may also be a complex scrubbing machine that is inserted into the pipe through a special opening. One of the longest metallic gas pipelines in the world is the Northern Lights pipeline, which is 5,470 kilometers long and links the West Siberian gas fields on the Arctic Circle with locations in Eastern Europe; in China, the recently completed "West Gas Supplying To East Project", yearly conveys 12 trillion cubic meters of natural gas from Xijiang Province gas fields to Beijing, the capital, in a 4,000 kilometer-long metallic pipeline.

The main differences of suggested Gas Transportation Method and Installation from modern metallic pipelines are:
1. The tubes are made from a lightweight flexible thin film (no steel or solid rigid hard material).
2. The gas pressure in tube equals an atmospheric pressure of 1- 2 atm. (Some current gas pipelines have pressure of 70 atm.).
3. Most of the filmic pipeline [except compressor (pumping) and driver stations] is located in the Earth-atmosphere at a high altitude (0.1-6 km) and does not have a rigid support (pillar, pylon, tower). All operating pipelines are located on the ground surface, underground or underwater.
4. The transported natural gas supports the air pipeline in the air above the route selected.
5. Additional aerial support may be made by employment of attached winged devices.
6. The natural gas pipeline can be used as an air transport system for oil and solid payloads with a maximum speed up of 250 m/sec.
7. The natural gas pipeline can be used as a transfer of a mechanical energy.

The suggested Method and Installation have remarkable cost-benefit advantages in comparison with all existing natural gas pipelines.

## Description of innovation

A gas and payload delivery gas/load pipeline is shown on figs. 1 - 5.

Fig.1 shows the gas/load delivery installation by air pipeline.

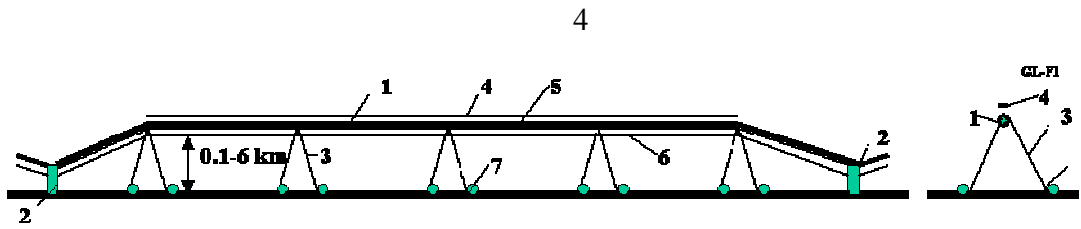

**Fig.1.** General view of aerial pipeline between two compression (pump) stations. (*a*) Side view, (*b*) front view. Notations: 1 - pipeline in atmosphere, 2 - compression station, 3 - tensile stress element, 4 - supportive wing device, 5 - night light, 6 - load (container) monorail, 7 - winch.

The installation works in the following manner: The compressor station pumps natural gas from storage into the tube (pipeline). The tube is made from light strong flexible gas-impermeable fireproof material (film), for example, from composed material (fibers, whiskers, nanotubes, etc.). A gas pressure is a less than an atmospheric pressure (up 1-2 atm). A natural (fuel) gas has methane as its main component with a specific density about 0.72 kg/m$^3$. Air has a specific density about 1.225 kg/cubic meter. That means that every cubic meter of gas (methane) or a gas mixture has a lift force approximately 0.5 kg. The linear (one meter) weight of a tube is less than a linear lift force of gas into the tube and the pipeline, therefore, has a lift force. The pipeline rises up and steadies at a given altitude (0.1 - 6 km), held fast by the tensile elements 3. The altitude of the aerial pipeline can be changed by the use of common winches 7. The compressor station is located on the ground surface and moves natural gas to the next compressor station that is ordinarily located at the distance 70 - 250 km from previous compressor station. Inside of the aerial pipeline there are valves (fig.4) dedicated to lock the tube tightly in case of it is punctured, ruptured or otherwise damaged. The pipeline has also the warning light indicator 5 for aircraft. The route selected for our example—see below—is well north of IATA-1, the new International Air Transport Association-approved flight-path for airliners coming from, or going to Europe via Hong Kong or Shanghai. Only international flights arriving or departing Beijing might come close to the selected example. Even if hit by an airliner, if the aircraft speed is greater than about 3% of the stress wave velocity, or greater than about 150 m/sec, the airplane's speed causes an immediate fracture that is independent of cable diameter, although the force on the vehicle's wing certainly is not! Fig.2 shows the gross-section of the gas pipeline and support ring. The light rigid tube ring keeps the lift force from gas tube, wing support devices, from monorail and load container.

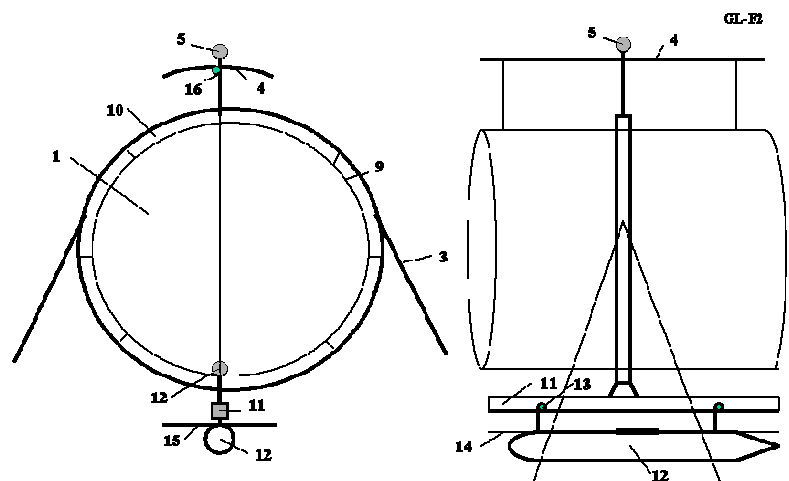

**Fig.2.** Gross-section and support ring of aerial natural gas pipeline. (*a*) front view, (*b*) side view. Notations: 9 - double casing, 10 - rigid ring, 11 - monorail, 12 - wing load container, 13 - rollers of a load container, 14 - thrust cable of container, 15 - container wing. Other notations are same fig.1.

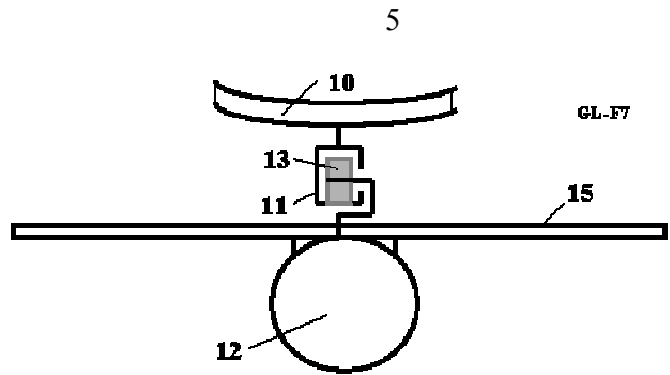

**Fig, 2a**. Front view of the winged load container, monorail, and suspension bracket. Other notations are same figs.1-2.

Fig.3 shows the compressor (pumping) station. The station is located on the ground and works in the following way. The engine 32 rotates compressor 31. That may be propellers located into rigid bogy connected to the flexible tubes of installation 1. The tubes are located at atmosphere. The propellers move the gas in given direction.

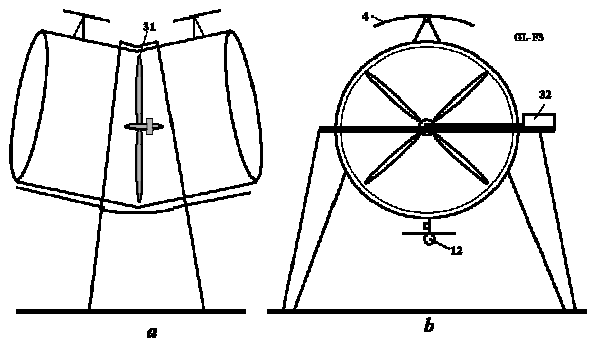

**Fig.3.** Compression (pump) station. Notations: 31 - compressor (propeller), 32 - engine. Other notations are same figs.1-3.

Fig.4 shows two variants a gas valve. The fist valve version is an inflatable boll. The ball fill out and closed the gas way. The second version is con conventional light flat choker in a form of circle.
The valve works in the following way. When the tube section is damaged, the pressure decreases in a section. The control devices measure it and, subsequently, the valves close the pipeline. The valve control devices have a connection with a central dispatcher, who can close or open any sub-section of the very long proposed natural gas pipeline.
Fig.4a shows the spherical valve (a ball) in a packed form. Fig.4b shows the spherical valve (a ball) in an open form.
The tubes of pipeline have a double wall (films) and gas streak between them. If the walls damage the streak gas flows out and the second film closes the hole in the first film and save a tube gas.

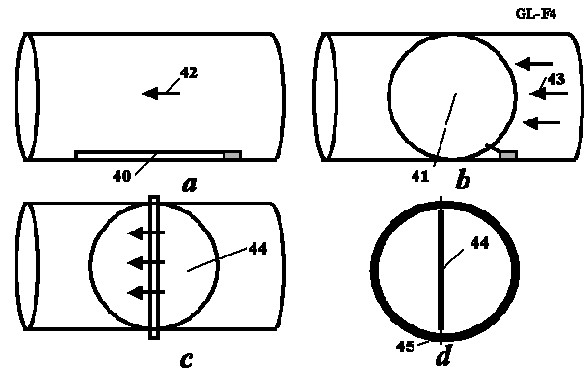

**Fig. 4**. Gas valve. (a)-(b) Inflatable valve, (c)-(d) Flat valve. Notations: 40 - inflatable valve in compact position, 41 - inflatable valve in fill out position, 42 - gas flow, 43 - gas pressure, 44 - flat choker.

The winged device 4 is a special automatic wing feature. When there is windy weather and the side wind produces a strong side force, the winged devices produce a strong lift force and support pipeline in fixed vertical position.

The winged device works in the following way: when there is a side wind the tube has the wing drag and the winged device creates needed additional lift force. All forces (lifts, drags, weights) are in equilibrium. The distance between the tensile elements 3 is such that the tube can resist the maximum storm wind speed. The system can have a compensation ring. The compensation ring includes ring, elastic element, and cover. The ring compensates the temperature's change of the tube and decreases a stress from a wind.

The suggested gas pipeline has big advantages over the conventional steel gas pipeline:
1. The suggested natural gas pipeline is to be made from a thin film that is hundreds of times less expensive than current gas pipelines currently made from steel tubes.
2. Construction time might be decreased from years to a few months.
3. There is no need to compress a gas, a huge saving of energy and expenditures for high-maintenance pumps.
4. No need for expensive ground surface and environmental damage during either the building or exploitation phases of the macroproject.
5. No environmental damage in case of pipeline's damage during use.
6. Easy to repair.
7. Decreasing energy for delivery.
8. Additional possibility of payload delivery in both directions.
9. If the aerial natural gas pipeline is situated at high altitude, it is more difficult for successful terrorist attacks and for gas thefts.
10. The suggested transportation system may be also used for a transfer of mechanical energy from one driver station to another.

More detail description of innovation the reader finds in publication. See [5]-[8].

Below, the authors have computed a macroproject suitable for Beijing region and the desert located in China's northwest territory. In addition, the authors have also solved additional problems, which appear in this and other macroprojects and which can appear as difficult as the proposed pipeline and transportation system itself. (The authors are prepared to discuss the technical problems with serious organizations that are interested in researching and developing these ideas and related macroprojects.)

## Methods of the estimation of the altitude gas pipeline

1. Gas delivery capability is
$$G = \pi D^2 V/4 \ [\text{m}^3/\text{sec}], \tag{1}$$
   where: $D$ is diameter of tube [m];
   $V$ is average gas speed [m/sec].

2. Increment pressure [n/m$^2$] is
$$P = \lambda L \rho V^2/2D, \tag{2}$$
   where: $\lambda$ is dimensionless factor depending on the wall roughness,
   the fluid properties, on the velocity, and pipe diameter ($\lambda = 0.01 - 0.06$);
   $L$ is the length [m];
   $\rho$ is fluid density [kg/m$^3$].

3. The dimensionless factor can be taken from graph [5, p.624]. It is
$$\lambda = f(R, \varepsilon), \tag{3}$$
   where: $R = VD\rho/\mu$ is Reynolds number, $\mu$ is fluid viscosity;
   $\varepsilon$ is measure of the absolute roughness of tube wall.

3. Lift force $F$ of the one meter length of pipeline is
$$F = (\upsilon_a - \upsilon_m)v, \tag{4}$$
   where: $\upsilon_a = 1.225$ kg/m$^3$ is air density for standard atmosphere;
   $\upsilon_m = 0.72$ kg/m$^3$ is methane density;
   $v = \pi D^2/4$ is volume of one meter length of pipeline.

4. Needed thickness $\delta$ of tube wall is
$$\delta = PD/2\sigma, \tag{5}$$
   where: $P$ is pressure [see Eq.(2)];
   $\sigma$ is safety stress. That is equals 100 - 200 kg/mm$^2$ for matter from current artificial fiber.

5. Weight of one meter pipeline is
$$W = \pi D \delta \gamma, \tag{6}$$
   where: $\gamma$ is specific weight of tube matter (film, cover). That equals about 0.9 - 2.2 kg/m$^3$ for matter from artificial fiber.

6. Air drag $D$ of pipeline from side wind is
$$D = C_d \rho V^2 S/2, \tag{7}$$
   where: $C_d$ is drag coefficient;
   $S$ is logistical pipeline area between tensile elements;
   $\rho$ is air density.

7. Needed power for delivery is
$$N = PG/\eta, \tag{8}$$
   where: $\eta \approx 0.9$ is a efficiency coefficient of a compressor station.

### Load transportation system under pipeline

1. Load delivery capability by wingless container is
$$G_p = kFV_p, \tag{9}$$
   where: $k$ is load coefficient ($k \approx 0.5 < 1$);
   $V_p$ is speed of container (load).

2. Friction force of wingless containers (rollers) is
$$F_c = fW_c, \tag{10}$$
   where: $f \approx 0.03 - 0.05$ is coefficient of roller friction;
   $W_c$ is weight of container between driver stations.

3. Air drug of container is

$$D_c = C_c \rho V^2 S_c / 2, \qquad (11)$$

where: $C_c$ is drug friction coefficient related to $S_c$;
   $S_c$ is cross-section area of container.

4. The lift force of a wing container is

$$L_c = C_L q S_{cw} = C_L \frac{\rho_a V^2}{2} S_{cw}, \qquad (12)$$

where: $C_L \approx 1 \div 1.5$ is coefficient of lift force; $q = \rho_a V^2 / 2$ is air dynamic pressure, N/m$^2$;
   $S_{cw}$ is wing area of container, m$^2$.

5. The drug of wing container may be computed by equation

$$D_C = C_L q S_{cw} / K = C_L \frac{\rho_a V^2}{2K} S_{cw}, \quad \text{where} \quad K = \frac{C_L}{C_D}, \qquad (13)$$

where $K \approx 10 \div 20$ is coefficient of aerodynamic efficiency; $C_D$ is air drag coefficient of wing container. If lift force of wing container equals the container weight, the friction force $F$ is absent and not necessary in monorail.

6. The delivery (load) capacity of the wing container is

$$G_c = \frac{W_1 V_c T}{d}, \qquad (14)$$

where $W_1$ is weight of one container, kg; $V_c = 30 \div 200$ m/s is container speed, m/s; $T$ is time, s; $d$ is distance between two containers, m.

7. The lift and drag of the wing device may be computed by Equations (12)-(13). The power needs for transportation system of wing container is

$$P_c = \frac{gWV_c}{K_c}, \qquad (15)$$

where $W$ is total weight of containers, kg; $g = 9.81$ m/s$^2$ if Earth gravity; $K_c \approx 10 \div 20$ is aerodynamic efficiency coefficient of container and thrust cable;

8. The stability of the pipeline against a side storm wind may be estimated by inequality

$$A = \frac{L_T + L_d - gW_T - gW_S}{D_T + L_d / K_d} > \tan \alpha > 0, \qquad (16)$$

where $L_T$ is lift force of given part of pipe line (conventionally it is distance between the tensile element, N; $L_d$ is lift force of wing device, N; $W_T$ is a weight of pipeline of given part, kg; $W_s$ is weight of the given part suspending system (containers, monorail, thrust cable, tensile element, rigid ring, etc.), kg; $D_T$ is drag of the given part of pipeline, N; $L_d$ is the lift force of the wing device, N; $K_d$ is an aerodynamic efficiency coefficient of wing device; $\alpha$ is angle between tensile element and ground surface.

## China Gas/Water Aerial Pipeline Macroproject

(*Tube diameter equals D = 10 m, gas pipeline has the suspension load transport system, the project is suitable for Beijing region –Gobi Desert*)

Let us take the distance between the compressor-driver stations 100 km and a gas speed $V = 10$ m/sec.

Gas delivery capacity is (Eq. (1))

$$G = \pi D^2 V / 4 = 800 \text{ m}^3/\text{s} = 24 \text{ billions m}^3 \text{ per year}.$$

For the Reynolds number $R = 10^7$ value $\lambda$ is 0.015, $P = 0.18$ atm (Eq. (2)-(3)). We can take V = 20 m/s and decrease delivery capacity by two (or more) times.

Lift force (Eq.4) of one meter pipeline's length equals $F = 39$ kg.

We take the thickness of wall as 0.15 mm for $\sigma = 200$ kg/sq. mm.

The cover weight of one-meter pipeline's length is 7 kg. The needed power of the compressor station (located at distance 100 km) equals $N = 10{,}890$ kW for $\eta = 0.9$.

**Load transportation System**

Let us take the speed of delivery equals $V = 30$ m/sec, payload capability is 20 - 25 kg per one meter of pipeline in one direction. Then the delivery capability for non-wing containers is 750 kg/s or 23 millions tons per year.

That is more than gas delivery (18 million tons per year). The total load weight suspended under the pipeline of length $L = 100$ km equals 2500 tons. If a friction coefficient is $f = 0.03$, the needed trust is 75 tons and needed power from only a friction roller drug is $N_1 = 22{,}500$ kW (Eq. (10)).

If the air drag coefficient $C_d = 0.1$, cross section container area $S_c = 0.2$ m$^2$, the air drag of a one container equals $D_{c1} = 2.2$ kg, and total drag 20,000 container of length 100 km is $D_c = 44$ tons. The need driver power is $N_2 = 13{,}200$ kW. The total power of transportation system is $N = 22500 + 13200 = 35{,}700$ kW. The total trust force is $77 + 44 = 121$ tons.

If $\sigma = 200$ kg/sq. mm, the cable diameter equals 30 mm.

The suggested delivery system can delivery a weight units (non-wing container) up to 100 kg if a selected length of container is 5 - 7.5 m.

The pipeline and container delivery capability may be increased at tens of times if winged containers are utilized. In this case we are not limited in load capability. Winged container needs a very lightweight monorail (or without it) and only in closed-loop thrust cable. That can be used for delivery water, oil or payload in containers. For example, if our system deliveries 4 m$^3$/second that is equivalent of a normal river (or a water irrigation canal) having a cross-section area equal to 20×2 m and water flowing speed 0.1 m/s. In other words, northwest China's planted desert dust suppression macroproject—the Great Green Wall [9]—can be fostered by delivery of irrigation water to the vegetation that may become available in AD 2008, just as the Olympic Games are played in Beijing, from the East Route of the "South-North Water Transfer Scheme" [10].

This particular macroproject system can transfer mechanical energy, we can transfer 35,700 kW for the cable speed at 30 m/sec, and 8 times more by the same cable having a speed 250 m/sec.

If the $\alpha < 60^\circ$ and wing of winged device has a width of 6 m, the system is stable against a side-thrusting storm wind of 30 - 40 m/second.